\documentclass[prl,twocolumn,showpacs,superscriptaddress,nofootinbib,10pt]{revtex4-2}

\pdfoutput=1
\usepackage[utf8]{inputenc}
\usepackage{amsmath,amsthm,amssymb,amsfonts}
\usepackage{mathtools}
\usepackage{graphicx}
\usepackage{subfigure}
\usepackage[colorlinks = true,
            linkcolor = blue,
            urlcolor  = blue,
            citecolor = blue,
            anchorcolor = blue]{hyperref}
\usepackage{mathrsfs}
\usepackage{bbold}
\usepackage[]{units}
\usepackage{bm}
\usepackage{braket}
\usepackage{color}
\usepackage{makecell}
\usepackage{enumitem}
\usepackage{upgreek}
\usepackage{blindtext}
\usepackage{graphics}
\usepackage{verbatim} 
\usepackage{adjustbox}

\usepackage[dvipsnames]{xcolor}
\usepackage{bbm}
\usepackage{longtable}
\usepackage[bb=boondox]{mathalfa}
\usepackage{array}
\usepackage{multirow}
\usepackage{tabularx}
\usepackage{float}
\usepackage{appendix}
\usepackage{placeins}
\usepackage{soul}

\usepackage{tikz}
\usetikzlibrary{shapes, arrows, patterns}

\newcommand{\be}{\begin{equation}}
\newcommand{\ee}{\end{equation}}
\newcommand{\bea}{\begin{eqnarray}}
\newcommand{\eea}{\end{eqnarray}}

\def\squareforqed{\hbox{\rlap{$\sqcap$}$\sqcup$}}
\def\qed{\ifmmode\squareforqed\else{\unskip\nobreak\hfil
\penalty50\hskip1em\null\nobreak\hfil\squareforqed
\parfillskip=0pt\finalhyphendemerits=0\endgraf}\fi}
\def\endenv{\ifmmode\;\else{\unskip\nobreak\hfil
\penalty50\hskip1em\null\nobreak\hfil\;
\parfillskip=0pt\finalhyphendemerits=0\endgraf}\fi}

\newcommand{\I}{\mathbbm{1}}

\makeatletter
\newtheorem*{rep@theorem}{\rep@title}
\newcommand{\newreptheorem}[2]{%
\newenvironment{rep#1}[1]{%
 \def\rep@title{#2 \ref{##1}}%
 \begin{rep@theorem}}%
 {\end{rep@theorem}}}
\makeatother

\newreptheorem{theorem}{Theorem}

\usepackage{graphicx}
\usepackage{braket}
\usepackage{amsmath,amsfonts,amssymb,amstext,amscd,amsthm,bbold}
\usepackage{comment,float}

\usepackage[dvipsnames]{xcolor}

\usepackage{lipsum}
\usepackage{dcolumn}
\usepackage{bm}
\usepackage{verbatim}

\usepackage{orcidlink}
\hypersetup{
	colorlinks=true,
	citecolor= BrickRed,
	linkcolor=BrickRed,
	filecolor=BrickRed,      
	urlcolor=BrickRed,
	pdftitle={Overleaf Example},
	pdfpagemode=FullScreen,
}

\usepackage{titlesec}
\titleformat{\section}
{\sffamily\bfseries} 
{\thesection} 
{1em} 
{} 

\titleformat{\subsection}
{\sffamily\bfseries} 
{\thesubsection} 
{1em} 
{} 

\usepackage[T1]{fontenc}
\usepackage[utf8]{inputenc}

\titlespacing\section{0pt}{12pt plus 4pt minus 2pt}{2pt plus 2pt minus 2pt}
\titlespacing\subsection{0pt}{12pt plus 4pt minus 2pt}{2pt plus 2pt minus 2pt}
\titlespacing\subsubsection{0pt}{12pt plus 4pt minus 2pt}{2pt plus 2pt minus 2pt}
\usepackage{lipsum} 

\makeatletter
\let\@afterindenttrue\@afterindentfalse
\makeatother

\begin{document}

\title{Certified Random Number Generation using Quantum Computers}

\author{Pingal Pratyush Nath}
\email[]{pingalnath@iisc.ac.in}
\affiliation{Indian Institute of Science,  C. V. Raman Road, Bengaluru, Karnataka 560012, India}

\author{Aninda Sinha }
\email[]{asinha@iisc.ac.in}
\affiliation{Indian Institute of Science,  C. V. Raman Road, Bengaluru, Karnataka 560012, India}
\affiliation{Department of Physics and Astronomy, University of Calgary, Alberta T2N 1N4, Canada}

\author{Urbasi Sinha }
\email[]{usinha@rri.res.in}
\affiliation{Raman Research Institute, C. V. Raman Avenue, Sadashivanagar, Bengaluru, Karnataka 560080, India}
\affiliation{Department of Physics and Astronomy, University of Calgary, Alberta T2N 1N4, Canada}

\begin{abstract}

In recent decades, quantum technologies have made significant strides toward achieving quantum utility. However, practical applications are hindered by challenges related to scaling the number of qubits and the depth of circuits. In this paper, we investigate how current quantum computers can be leveraged for practical applications, particularly in generating secure random numbers certified by Quantum Mechanics. While random numbers can be generated and certified in a device-independent manner through the violation of Bell's inequality, this method requires significant spatial separation to satisfy the no-signaling condition, making it impractical for implementation on a single quantum computer. Instead, we employ temporal correlations to generate randomness by violating the Leggett-Garg inequality, which relies on the No-Signaling in Time condition to certify randomness, thus overcoming spatial constraints. By applying this protocol to existing quantum computers, we demonstrate the feasibility of secure, semi-device-independent random number generation using low-depth circuits with single-qubit gates.

\end{abstract}

\maketitle
\section{Introduction}

Over the past couple of years, significant efforts have been dedicated to the development of quantum technologies, in a race to attain quantum supremacy \cite{arute2019quantum, wu2021strong, zhong2020quantum}. Recent research \cite{kim2023evidence} has demonstrated that quantum computers have entered the era of quantum utility, where they can perform reliable computations on a scale that exceeds classical brute-force methods, offering exact solutions to complex computational problems. Despite this milestone, the utilization of quantum computers for practical advantages remains a distant aspiration, primarily due to challenges in scaling the number of qubits \cite{preskill2018quantum}. 
Diligent efforts are ongoing to realize this dream in the earliest possible way, with various efforts focused on leveraging the limited available resources to perform classically challenging tasks \cite{bharti2022noisy,cerezo2021variational,endo2021hybrid}.

However, with progressively improved control over noise in state-of-the-art quantum computers \cite{google2023suppressing,acharya2408quantum,murali2020software,corcoles2019challenges,urbanek2021mitigating,giurgica2020digital,kandala2019error,ball2021software,nation2021scalable,ferracin2024efficiently,sharma2020noise,suzuki2022quantum,cincio2021machine,maciejewski2020mitigation,lowe2021unified,he2020zero,nachman2020unfolding,aghaee2025scaling}, practical quantum advantages can be realized using the resources offered by quantum mechanics. This paper demonstrates for the first time one such practical advantage of current quantum computers, utilizing only a single qubit, to generate secure random numbers that are certified by the principles of quantum mechanics.

Randomness generation \cite{marsaglia1990toward,marsaglia1991new,marsaglia2003random,l2012random,hull1962random,jennewein2000fast,stipvcevic2014true,hellekalek1998good} plays a crucial role in various domains, including Cryptography, Statistics, and Biology, with applications ranging from encryption key generation to simulating complex systems and even in gaming. Conventionally, computers generate random numbers using mathematical algorithms that rely on an initial random seed. These deterministic processes, known as Pseudo Random Number Generators (PRNG) \cite{blum1986simple,vazirani1984efficient}, are limited by their predictability, as their randomness is entirely dependent on the initial seed. Consequently, PRNGs are unsuitable for applications requiring high-security standards.

In contrast, True Random Number Generators (TRNGs) \cite{stipvcevic2014true,yu2019survey,fischer2002true,bagini1999design,sunar2006provably} utilize physical processes such as atmospheric noise or radioactive decay, which are inherently non-deterministic. This approach provides a high degree of entropy, essential for generating cryptographic keys that are resistant to guessing or brute-force attacks. Cryptographic algorithms heavily depend on the secrecy of distributing cryptographic keys, necessitating the use of random numbers as seeds that cannot be predicted by potential eavesdroppers.

However, trusting the manufacturer of a TRNG is paramount to ensuring the integrity of the generated random numbers. A potential security threat is the memory stick attack \cite{acin2016certified} 
, where high-quality random numbers are stored in a memory stick within the TRNG device, posing a risk to security. While statistical tests \cite{rukhin2001statistical,bassham2010sp,bassham2010statistical} can assess the uniformity of generated bits, certifying the randomness of the source remains a challenging problem. Moreover, characterizing the quality of the random bits or the entropy of the source based on the generated outputs is a complex task. Another challenge with a TRNG is that it is a physical device and, like all hardware, it degrades over time.

Quantum processes due to their inherent randomness are excellent sources for generating random numbers \cite{herrero2017quantum, ma2016quantum}. Quantum correlations violate certain inequalities which cannot be violated by classical correlations. A class of these constraints known as Bell inequalities \cite{bell1964einstein,brunner2014bell,cirel1980quantum,franson1989bell,peres1999all,aspect1999bell} can be used to certify the quantum nature of the random bits generated \cite{acin2016certified} in a device independent way from just the statistics of the measurement outcomes without any assumptions on the device used. This novel idea of generating device-independent randomness certified by quantum mechanics was first demonstrated by violating the CHSH inequality \cite{pironio2010random}, which was followed by loophole-free demonstrations of the Bell inequality violation experiment \cite{shalm2021device, bierhorst2018experimentally, liu2018device, zhang2020experimental, liu2018high, shen2018randomness, abellan2015generation}. However, in order to generate device-independent random numbers using this protocol, the two stations where the entangled pair is measured needs to be spatially separated by a large distance to ensure no-signaling. This large separation makes it extremely hard to produce consumer-grade device-independent random number generators \cite{pironio2018certainty}. Moreover, this requirement for spatial separation hinders the implementation of this protocol on a single superconducting quantum chip.

The temporal analogue of the Bell Inequalities, viz. the Leggett-Garg Inequalities \cite{emary2013leggett, leggett1985quantum}, can be used for certifying quantum randomness in a table-top experiment \cite{joarder2022loophole}. This was demonstrated in a photonic setup \cite{nath2024single} where random numbers were generated in a loophole free experiment for LGI violation. The experiment involves single-photon sources, interferometers for generating temporal correlations and projective measurements for randomness generation. Overcoming the distance barrier seen in Bell experiments, this approach presents a promising avenue for practical implementation.  While this opens up a new paradigm for randomness generation, the implementation itself requires a large amount of optical and optomechanical infrastructure which is dedicated to the cause. A significant step forward would be to use the developed methodology on commercially available devices that need not be custom-made for the purpose. This brings us to a question: Can we use for instance a NISQ quantum computer to generate such random numbers by violating LGI? Not only will this be a fantastic practical use case for the current quantum computers, but it will in fact be a very unique platform that brings forth the use of a quantum computer in a niche quantum security application. 

In this paper, we go on to do just that successfully! We adopt this protocol, to generate random numbers on available IBM superconducting quantum computers \cite{qiskit2024}. Utilizing low-depth circuits with simple one-qubit gates and projective measurements, this approach provides a feasible and secure method for practical semi-device-independent random number generation. Although cloud-based quantum computers were used previously to generate random numbers \cite{li2021quantum, jacak2021quantum, orts2023quantum, kumar2022quantum, sinha2023programmable}, their quantum nature cannot be certified device-independently, making them less secure. In contrast, our implementation leverages Leggett-Garg Inequality (LGI) violation to certify the randomness coming from a quantum source, thus offering a practical use case for NISQ devices.

\section{Protocol for Randomness Generation}
The Leggett Garg Inequality (LGI) characterizes a single-time evolving system where measurements of a dichotomic variable $Q$ with eigenvalues $+1$ and $-1$ are taken at different times. The inequality is expressed as:

\begin{equation}\label{eq: LGI Inequality}
\braket{Q_1Q_2} + \braket{Q_2Q_3} - \braket{Q_1Q_3} \leq 1 .
\end{equation}

Here, $Q_i = Q(t_i)$ represents the measurement outcome at time $t_i$ in a time sequence $t_1<t_2<t_3$. The correlation functions are defined as:

\begin{equation}\label{eq: correlation functions}
\braket{Q_{i}Q_{j}} = \sum_{a_{i},a_{j} = \pm1}a_{i}a_{j}P(a_i, a_j|Q_i, Q_j) ,
\end{equation}

where $P(a_i,a_j|Q_i, Q_j)$ denotes the probability of obtaining outcomes $a_i$ and $a_j$ for $Q_i$ and $Q_j$ respectively. The quantum mechanical violation of this inequality, capped at $1.5$, is associated with the breach of assumptions defining macrorealism \cite{emary2013leggett,leggett1985quantum,mal2016temporal,nath2024single}.

LGI can be derived from Predictability and No Signaling in Time (NSIT) \cite{kofler2008conditions,clemente2015necessary,kofler2013condition}, similar to the derivation of Bell-CHSH inequality from Predictability and No Signaling across spatial separation \cite{mal2016temporal, cavalcanti2012bell,halliwell2016leggett}. In the Bell Scenario, if the measurement outcomes of an entangled state at two well-separated measurement stations violate the Bell Inequality, they are confirmed to be random \cite{pironio2010random,pironio2018certainty,acin2016certified,cavalcanti2012bell}. Similarly, if an experiment's measurements adhere to the constraints of the NSIT condition while violating LGI, the measurement outcomes are random according to the predictability condition.

For the three-time LGI, the No Signaling in Time conditions are:
\begin{align}\label{eq: NSIT}
& P(+|Q_2) = P(++|Q_1, Q_2) + P(-+|Q_1,Q_2) \nonumber\\
& P(+|Q_3) = P(++|Q_1, Q_3) + P(-+|Q_1,Q_3) \nonumber\\
& P(+|Q_3) = P(++|Q_2, Q_3) + P(-+|Q_2,Q_3)
\end{align}

The Leggett-Garg inequality places limitations on predicting a system's future behavior based on its past, whereas the no-signaling-in-time criterion asserts that a system's future behavior should be independent of earlier measurements. When both conditions are met, a measurement can produce an unpredictable outcome. This unpredictability is valuable in security applications, such as cryptographic protocols that require a source of secure randomness. A test can be formulated to confirm the quantum nature of these random numbers, utilizing the protocol to design an experiment satisfying NSIT and violating LGI, certifying random outputs according to Quantum Mechanics.

\begin{figure}
    \centering
    \includegraphics[width = 0.3\textwidth]{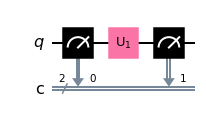}
    \includegraphics[width = 0.4\textwidth]{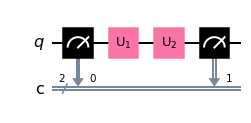}
    \includegraphics[width = 0.4\textwidth]{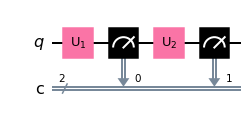}
    \caption{The circuits for different measurement setting $t_1t_2$, $t_1t_3$ and $t_2t_3$. $U_1$ and $U_2$ denotes the rotation operators with angles $\theta_1$ and $\theta_2$}
    \label{fig: circuits}
\end{figure}

We will adopt the analysis developed in \cite{nath2024single} to design the experiment for generating randomness. We will use a two-level system and perform projective measurements on it at different times, and the conditions for certifying randomness can be evaluated from the measurement outcomes. The amount of randomness can also be evaluated from the probabilities $P(a_i,a_j|Q_i,Q_j)$, 

\begin{align}\label{eq:minimumentropydefinition}
    H_{\infty}(AB|XY) &= -\log\{\text{max}_{a_i, a_j}P(a_i, a_j|Q_i, Q_j)\}\nonumber\\
                      &= -\text{min}_{a_i, a_j}\log\{P(a_i, a_j|Q_i, Q_j)\}.\nonumber\\
\end{align}

\noindent where $H_\infty$ is called Genuine Randomness \cite{acin2011randomness}. A minimum bound on the amount of randomness generated based on the LGI violation $I$ was given in \cite{nath2024single}, 
\begin{equation}\label{lowerboundstatesecure}
H_{\infty}(AB|XY) \geq -\log_2 \left(\frac{1 + \alpha + \sqrt{1-2\alpha}}{2}\right).
\end{equation}
with $\alpha$ defined as $I - 1$, where $I$ represents the LGI violation.

\section{Quantum Circuit} We employ a simplified circuit to generate random numbers by concurrently violating LGI and adhering to the NSIT constraints. The most general one qubit state, characterized by the parameters $n_x$, $n_y$, and $n_z$, is expressed as follows:
\be 
\rho = \frac12 (\I + \Vec{n}\cdot \Vec{\sigma}) , \ \Vec{n} =(n_x,n_y,n_z) \in \mathbbm{R}^3
\ee 
such that $n_x^2+n^2_y+n^2_z \leqslant 1$. To keep things simple we set the parameters as $n_x=0$, $n_y=1$, and $n_z = 0$, which corresponds to the state, $(\ket{0}- i\ket{1})/\sqrt{2}$. For the time evolution, we opt for the basic rotation gates $U_1$ and $U_2$ parameterized by angle $\theta$ as, 

\be 
U_i = 
\begin{pmatrix}
 \cos[\theta_i] &  \sin [\theta_i] \\
 -\sin [\theta_i] &  \cos [\theta_i]
\end{pmatrix} 
\text{ for } i=1,2 , \quad \theta_i \in \mathbbm{R} 
\ee 
We perform projective measurements at time instances $t_1$, $t_2$, and $t_3$ in the computational basis. The projectors for this basis are defined as follows:
\be 
P_+ = \begin{pmatrix}
1 & 0 \\
0 & 0
\end{pmatrix}, \ 
P_- =
\begin{pmatrix}
0 & 0 \\
0 & 1
\end{pmatrix} .
\ee 

It is important to note that adopting a different measurement basis would necessitate additional gates, introducing potential sources of errors. 
 Using this initial state and the form of unitaries and measurements mentioned above we compute the expressions of LGI and the NSIT conditions 
and find the values of $\theta_1$ and $\theta_2$ for different LGI violations as shown in table \ref{table:pure_state_params_table}.



\begin{table}
\centering
\renewcommand{\arraystretch}{1.3} 
\setlength{\tabcolsep}{12pt} 
\begin{tabular}{|c|c|c|}
\hline
\textbf{LGI} & \boldmath$\theta_1$ & \boldmath$\theta_2$ \\
\hline
1.05 & 267.061 & 142.144 \\
\hline
1.10 & 267.088 & 142.131 \\
\hline
1.15 & 267.117 & 142.116 \\
\hline
1.20 & 267.148 & 142.101 \\
\hline
1.25 & 267.182 & 142.084 \\
\hline
1.30 & 267.220 & 142.065 \\
\hline
1.35 & 267.263 & 142.043 \\
\hline
1.40 & 267.315 & 142.017 \\
\hline
1.45 & 267.384 & 141.983 \\
\hline
1.50 & -75.922 & -75.922 \\
\hline
\end{tabular}
\caption{The parameters $\theta_1$ and $\theta_2$ correspond to the rotation gates for time translations $t_1 \rightarrow t_2$ and $t_2 \rightarrow t_3$, based on the specified initial state and projective measurements. The circuits utilizing these $\theta$ values exhibit a violation of the LGI at a specific point while also satisfying all NSIT conditions, enabling secure randomness generation.}
\label{table:pure_state_params_table}
\end{table}

 In principle we can start with a different initial state, and choose more general measurements, which will lead to different parameters for the Unitaries. For example, as shown in Table \ref{table:mixed states} in the Appendix, we have demonstrated that starting with a mixed state allows for the design of an appropriate circuit. We emphasize that this choice of circuit for our algorithm might not be the most optimized choice and further research is warranted to solve the equations and identify the most efficient circuit for the algorithm. Regardless, the RNG does not depend on the choice of the circuit, only the complexity of implementing the algorithm will differ.
 
 It is important to note that the certification protocol here is semi-device independent because while deriving the bound for genuine randomness in Equation \ref{lowerboundstatesecure} it was assumed that the state of the system used is two-dimensional and the measurements at time $t_1$ and $t_2$ are projective measurements\cite{nath2024single}. The circuit we used above is one of the possible choices of the family of circuits given these constraints.

\section{IBMQ Results}

\begin{figure}
    \centering
    \includegraphics[width = 0.5\textwidth]{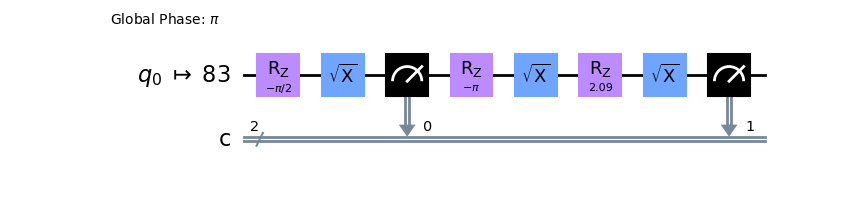}
    \includegraphics[width = 0.5\textwidth]{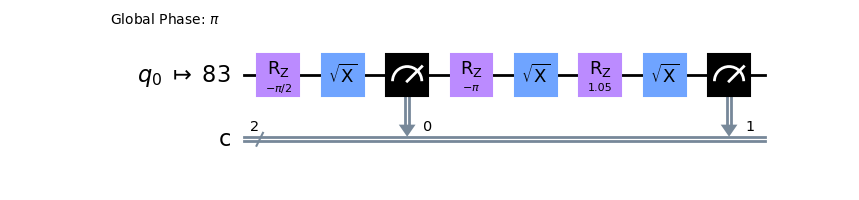}
    \includegraphics[width = 0.5\textwidth]{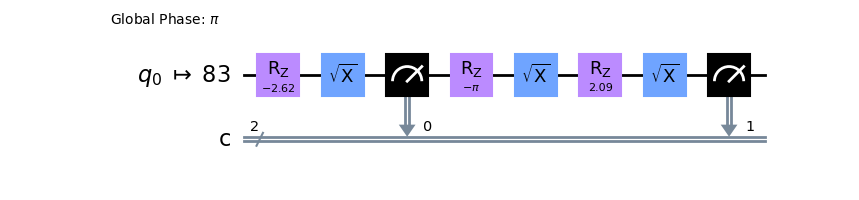}
    \caption{Circuits for correlation measurements of $t_1t_2$, $t_2t_3$ and $t_1t_3$ transpiled in the IBMQ Brussels backend. The Unitaries for rotation operators involving the angles $\theta_1$ and $\theta_2$ are decomposed in terms of the $R_Z$ and $SX$ gates available in the backend. The qubit 12 was selected after analyzing the best possible layout for our circuit using the mapomatic algorithm.}
    \label{fig: Circuit post transpilation}
\end{figure}

We utilized IBM Quantum Hardware for the generation of random numbers through the violation of the Leggett Garg Inequality. The unitaries in the circuits can easily be decomposed into a sequence of Z-rotation($R_Z$) and $SX$ gates, facilitating implementation in the hardware with minimal error rates. The circuits computing the correlations $\braket{Q_1Q_2}$, $\braket{Q_1Q_3}$, and $\braket{Q_2Q_3}$ after transpilation in the IBM backends can be decomposed into $SX$ Gates and $R_{Z}$ Gates as shown in Figure \ref{fig: Circuit post transpilation}.

\begin{figure}
    \centering
    \includegraphics[width = 0.5\textwidth]{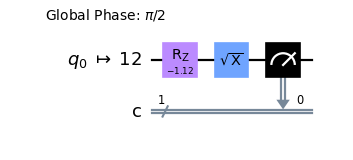}
    \includegraphics[width = 0.5\textwidth]{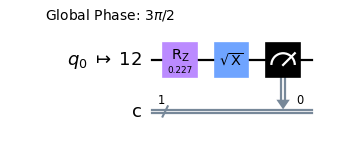}
    \caption{Circuit for computing the one time probabilities at $t_2$ and $t_3$. These circuits are utilized in the verification of the NSIT conditions and are not used in generating random bits. These circuits also can be decomposed in terms of the $SX$ and $R_Z$ gates followed by a single measurement. }
    \label{fig :NSIT circuits}
\end{figure}

To verify the No-Signaling In Time (NSIT) conditions, two additional circuits perform measurements solely at $t_2$ and $t_3$(Figure \ref{fig :NSIT circuits}) without prior measurements. The outcomes from these circuits, coupled with the results from correlation calculations, are employed to validate Eq \ref{eq: NSIT}. The concurrent violation of LGI and satisfaction of NSIT conditions collectively ensure the unpredictability of the outputs generated in the correlation measurements. The code \cite{githubrepo} for this work is available on \href{https://github.com/Pingal-Pratyush-Nath/Certified-Randomness-on-a-Quantum-Computer}{Github}.

\begin{figure}
    \centering
    \includegraphics[width=1\linewidth]{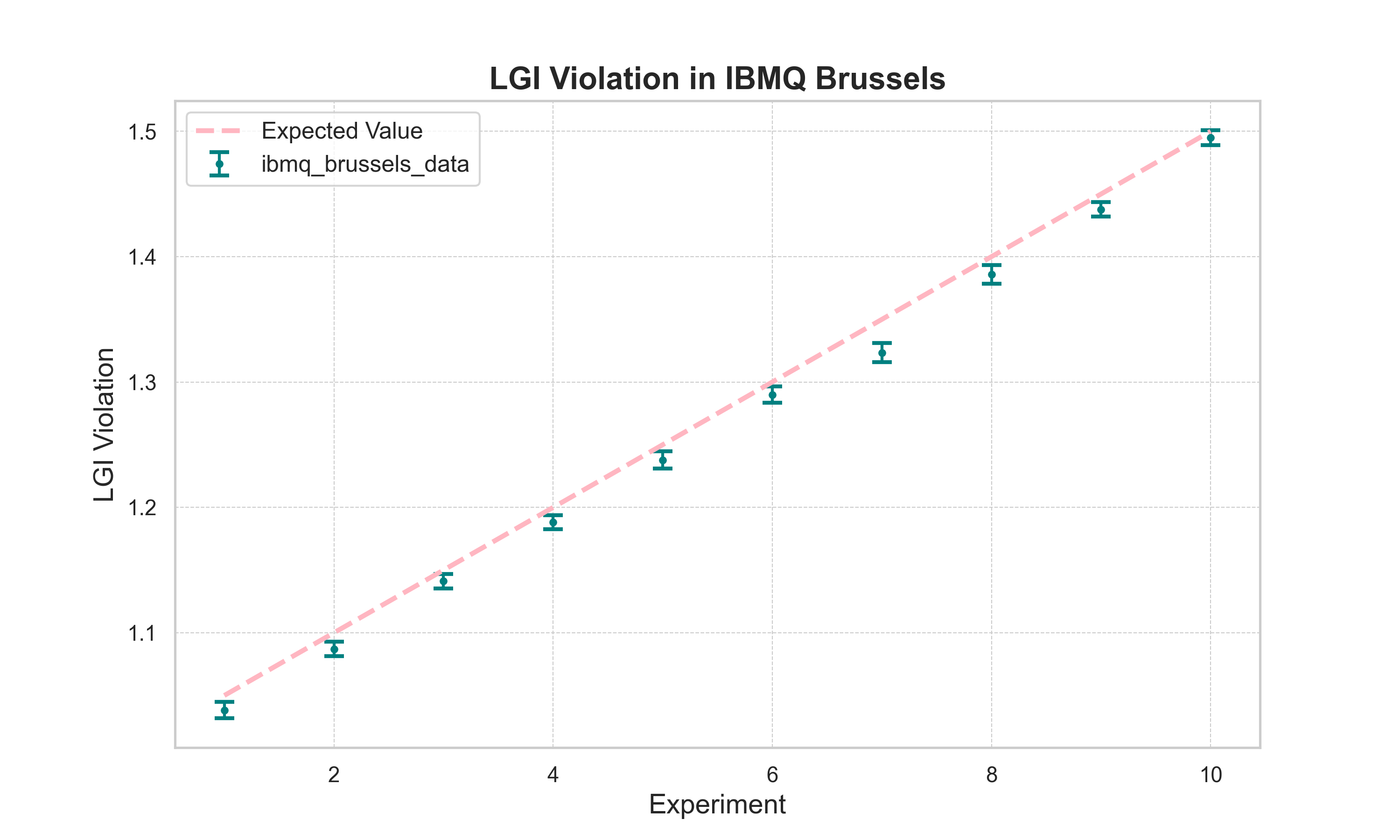}
    \caption{LGI violation experiment in IBMQ Brussels. We repeated the experiment for each value $10$ times, each of the experiment was run for $50,000$ shots.
    We observe that for all cases the experimental results are slightly lower than the expected values, which is due to the noise factors in the backend as demonstrated later.}
    \label{fig:lgi_violation_brussels}
\end{figure}

\begin{figure}
    \centering
    \includegraphics[width=1\linewidth]{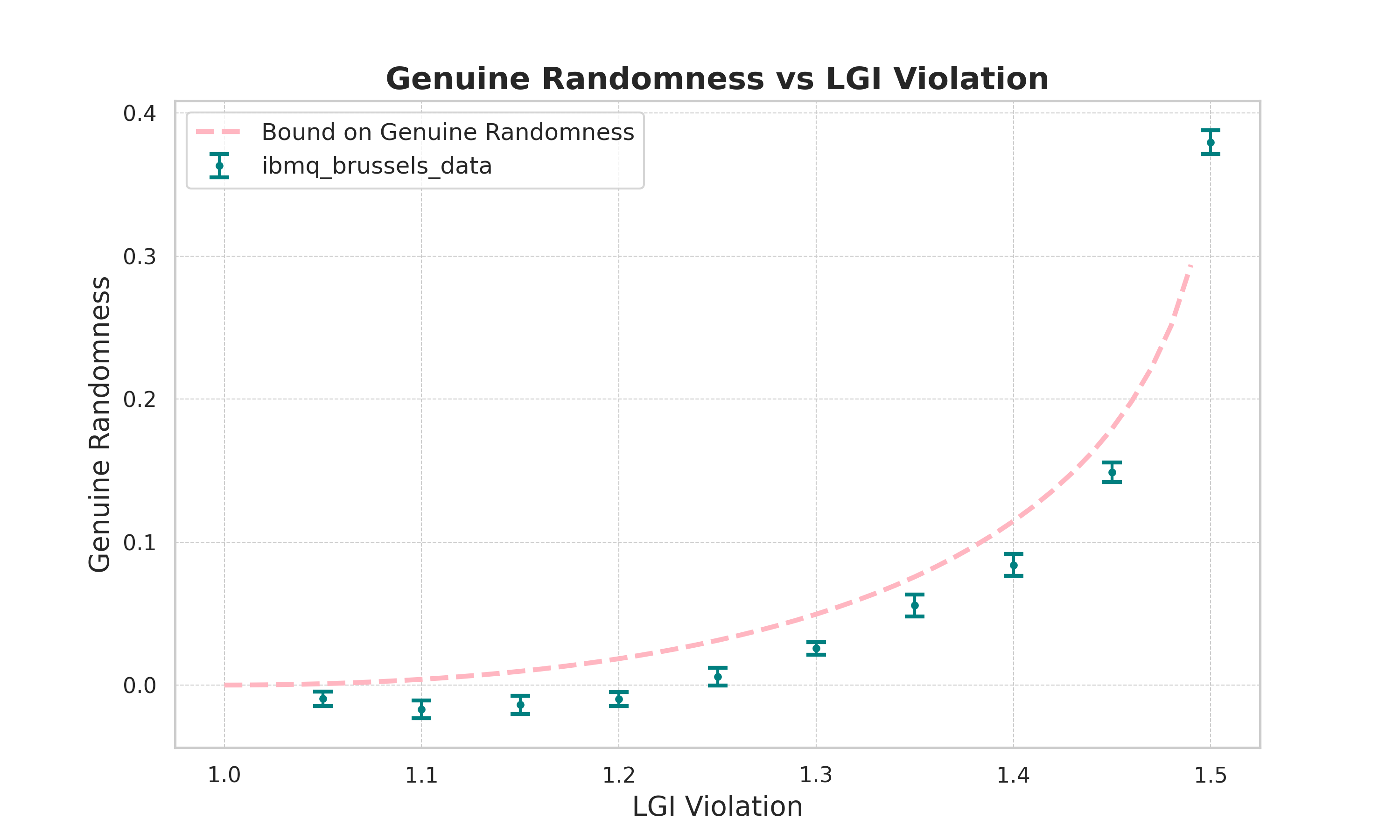}
    \caption{Genuine Randomness vs LGI violation plotted alongside the theoretical analytical bound for the experiment in IBMQ Brussels. The genuine randomness spread is a bit lower than the expected lower bound because the results of the LGI values in the experiment were lower than the expected values.}
    \label{fig:genuine_randomness_brussels}
\end{figure}

In each experiment we employ the five circuits for $N = 50,000$ shots each and compute the expected LGI and NSIT values. We repeat the experiment for each value of LGI violation $10$ times and see that the spread of LGI violation is around the range of the expected LGI value (Figure \ref{fig:lgi_violation_brussels}) and the NSIT conditions are satisfied up to an order $10^{-2}$. In each run of the experiment, we generate $2N$ bits from each of the first three sub-runs of the experiment for calculating the correlations. In order to protect the random bits from the attacks involved in state preparation we discard the first bit and employ conditional probabilities to compute the Genuine Randomness as shown in \cite{nath2024single}. The Genuine Randomness computed this way follows the bound {\bf of} given in Equation \ref{lowerboundstatesecure} as shown in Figure \ref{fig:genuine_randomness_brussels}.

\textit{Noise Mitigation} : In order to mitigate the noise in the quantum hardware we employed multiple techniques. We transpiled the original circuit against our backend to decompose it in terms of the available gates in the backend. We used mapomatic library\cite{nation2023suppressing} to select the layouts/ qubits in which our circuit fits. 
Then we used the mapomatic algorithm to score the best possible layout for our circuit in terms of the mapomatic score, which is calculated by combining the noise rates of each of the operations in the circuit for the noise parameters of the layout. Apart from the major experiment conducted in the IBMQ Brussels backend we also generated secure random numbers using some deprecated IBM backends: IBM \textit{Perth}, IBM \textit{Lagos}, and IBM \textit{Kyoto}. Certification was achieved through the successful violation of the Leggett-Garg Inequality and the satisfaction of the No Signaling in Time Conditions. 
The results of these experiments are there in the appendix.

\section{Noise Analysis}
\begin{figure}
    \centering
    \includegraphics[width=1\linewidth]{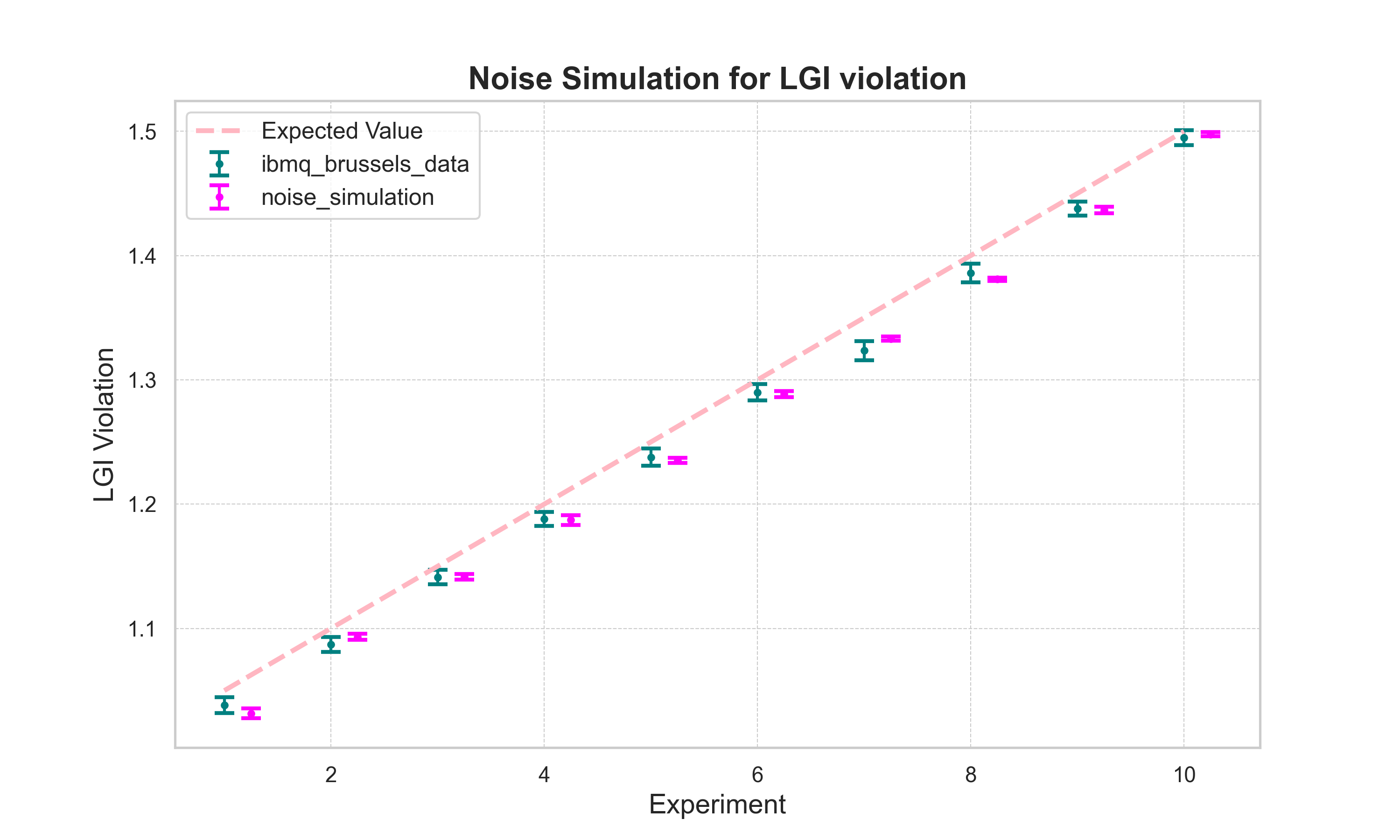}
    \caption{Noise simulation of the experiment using the noise parameters from the IBMQ Brussels backend, compared with the results of the actual experiment. Each experiment is conducted with $50,000$ shots and repeated $10$ times. We displaced the noisy simulation slightly from the actual one for better visibility. The close but not complete agreement between the simulation and experimental results demonstrates the impact of noise on the system.}
    \label{fig:lgi_violation_brussels_noise}
\end{figure}

In all of the above experiments we saw that the LGI value of the experiment is lower than the expected LGI value. To analyze the noise, we started with some sanity checks on the results. We ran the experiment in the qiskit Aer simulator(Appendix) and verified that it matches the exact result. We then imported the noise parameters from the device at the time of running the experiment and created a noise model from these noise parameters. Using this noise model on the Aer Simulator we ran the experiment and the results of this noisy simulation match with those of the original experiment as shown in Figure \ref{fig:lgi_violation_brussels_noise}. 
For better visibility of the actual results with the noise simulation, we displaced them slightly on the horizontal axis.

Although the experimental results are very close to the expected values, we want to address the potential sources of errors. The circuits used consist of $R_z$ and $SX$ gates. The $R_z$ gates are implemented flawlessly without any noise because they are diagonal gates, which can be implemented virtually in hardware through frame changes, resulting in zero error and no time duration. On the other hand, the $SX$ gates have an error rate of approximately $10^{-4}$. Although this error rate is very low compared to two-qubit gates such as CNOT and ECR(Echoed Cross Reasonance), it could still be a possible source of error.
\begin{figure}
    \centering
    \includegraphics[width=0.8\linewidth]{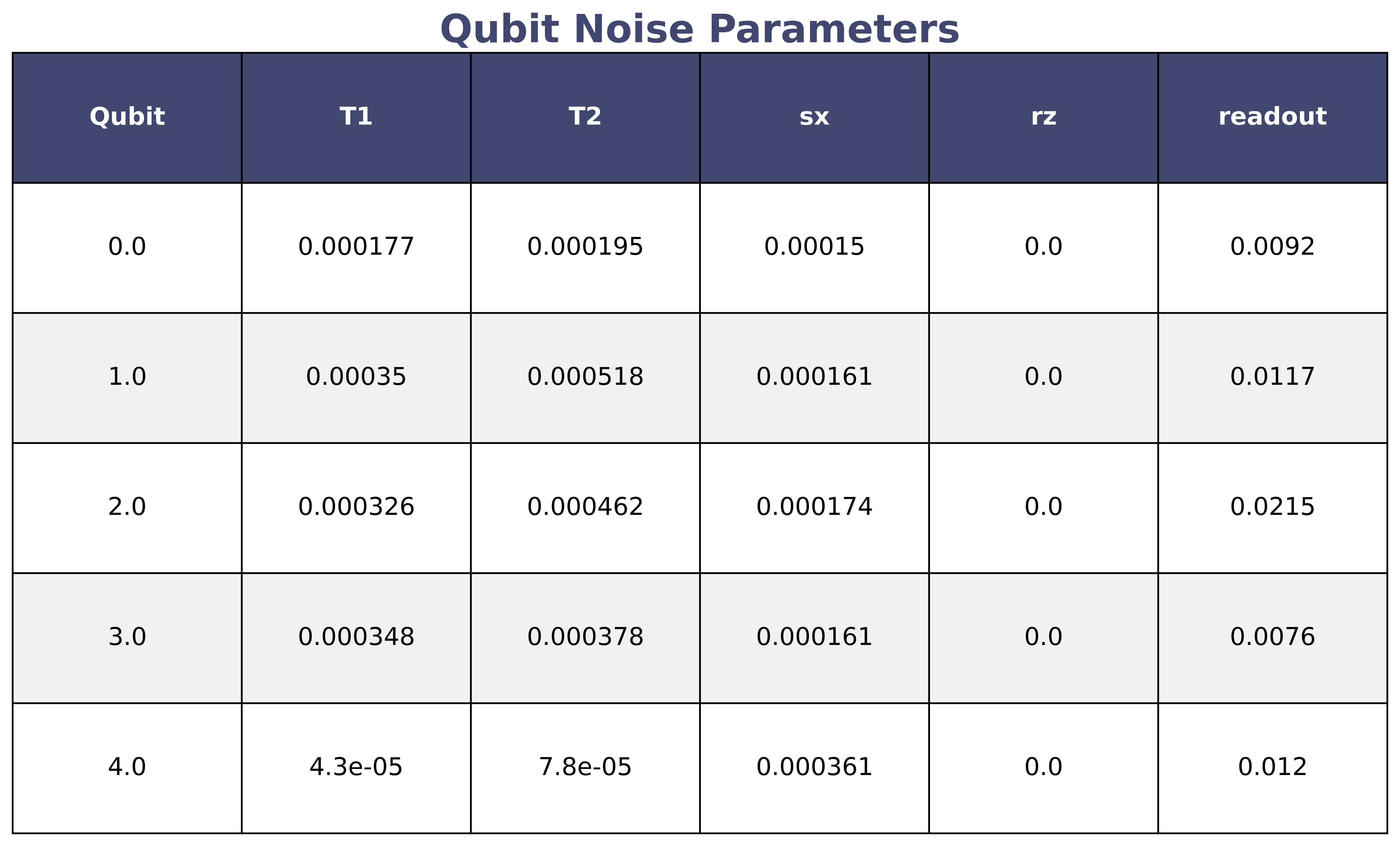}
    \caption{$T1$(Thermal relaxation time), $T2$(dephasing time), $SX$-error rates, $R_z$-error rates and readout error rates for randomnly selected qubits in the ibmq brussels backend}
    \label{fig:qubitnoiseparameters}
\end{figure}

The readout errors are significant, compared to the gate error rates. The readout error rates and gate error rates for selected qubits in the IBMQ Brussels backend are shown in Figure \ref{fig:qubitnoiseparameters}

\begin{figure}
    \centering
    \includegraphics[width=1\linewidth]{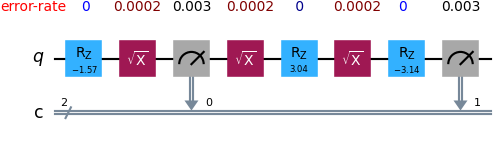}
    \caption{Error rates of the gates in the one-qubit circuit used to compute two-time correlations. The error analysis includes the gate error rates for $R_z$ and $SX$, as well as measurement errors. The $R_z$ gates are implemented flawlessly, with no detectable error. The $SX$ gates exhibit minimal errors, contributing only slightly to the overall noise. Despite the generally high readout errors, qubits with the lowest readout error rates were carefully selected to ensure the most accurate measurements possible.}
    \label{fig:errorrate}
\end{figure}

\begin{figure}
    \centering
    \includegraphics[width=1\linewidth]{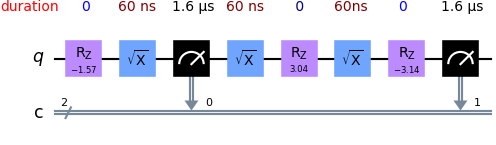}
    \caption{Duration of the elements in the one-qubit circuit to compute the two time correlations. The $R_z$ gates are implemented instantaneously with no measurable duration. The $SX$ gates operate on the scale of nanoseconds, while the measurement process occurs on the scale of picoseconds.}
    \label{fig:duration}
\end{figure}

Regarding decoherence errors, we computed the total time required to run the circuit by calculating the implementation time for each element, as shown in Figure \ref{fig:duration}. The $R_z$ gates are implemented instantly, while the $SX$ gates require time on the order of nanoseconds. The measurements take more time, on the order of microseconds. 
 Consequently, the entire circuit is executed in a few microseconds. Given that the decoherence times for the qubits in our backend are on the order of $10^{-4}$ seconds, the circuit is safely implemented within the decoherence time.

The $T1$ (thermal relaxation time) and $T2$(dephasing time) for some qubits are shown in Figure \ref{fig:errorrate}. This analysis demonstrates that our algorithm is well suited for implementation on the best available qubits in the back-end without suffering from decoherence.

\section{Advantages over the two particle scenario: } 
\begin{figure}
    \centering
    \includegraphics[width=1\linewidth]{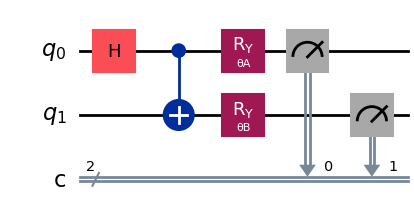}
    \caption{Circuit for violating the Bell inequality. A maximally entangled state is created and distributed between Alice and Bob. The measurement settings \(\theta_A\) and \(\theta_B\) are randomly selected from the possible choices.}
    \label{fig:bell}
\end{figure}

We demonstrate that it is possible to violate Bell's inequality using a quantum computer. This can be achieved by creating a maximally entangled state and selecting specific measurement bases for each qubit. In our example, we chose the measurement angles for Alice as $\theta_a = 0$ and $\theta_a^{\prime} = \pi/4$, and for Bob as $\theta_b = \pi/8$ and $\theta_b^{\prime} = 3\pi/8$, resulting in a Bell violation of $2\sqrt{2}$. For each iteration, a random seed was used to select measurement settings for Alice and Bob, and the outcomes was then used to compute the correlations. The corresponding quantum circuit for this experiment is shown in Figure \ref{fig:bell}.

However, the bits generated from the measurement outcomes of Alice and Bob in the above experiment cannot be certified as generating certified randomness from Bell inequality violations requires the additional constraint of satisfying the No-Signaling condition. To meet this requirement, the two measurement stations (Alice and Bob) must be sufficiently separated so that Alice is unaware of Bob's random seed and vice versa. Currently, this level of separation cannot be achieved, as communication between quantum computers is not feasible. Nevertheless, our protocol satisfies the necessary conditions for certified randomness, as the circuits are designed to violate the Leggett-Garg inequality (LGI) while also fulfilling the No-Signaling-in-Time condition.

Although the present form of the protocol is not entirely loophole-free, it offers a viable method for generating certified random numbers using quantum computers, something that is not possible in the Bell scenario, as demonstrated above.

 \section{Loopholes}

 We briefly address the potential loopholes in our experiment. For the clumsiness loophole \cite{huffman2017violation, wilde2012addressing} our experiment was designed in such a way that a measurement at an earlier time cannot signal to a measurement performed later. This was ensured by setting the parameters for the unitaries in such a way that they satisfy the No Signaling in Time condition, which is a necessary condition for the measurements to be noninvasive \cite{emary2017ambiguous}. The results from our experiment which satisfy the NSIT condition upto a tolerance of $10^{-2}$ further verify our results.  The detection efficiency loophole and the multi-photon emission loophole are irrelevant for LGI (Leggett-Garg Inequality) violation on superconducting quantum computers. Additionally, the coincidence loophole does not apply since our experiment does not involve heralding qubits. The preparation state loophole is automatically closed by the state preparation procedures of the IBM quantum chips, as they consistently produce the same initial state. We leave a fully loophole free implementation for future work.

\section{Noise Mitigation using Mthree}

We used IBM error mitigation techniques \cite{nation2021scalable} to further reduce readout errors in our experiment. The primary motivation for this approach was to strengthen the NSIT condition by eliminating classical sources of errors, particularly readout errors. Among the various sources of errors, measurement errors were the most dominant as in fig.(\ref{fig:qubitnoiseparameters}), and their careful mitigation is crucial to obtain more accurate values for Leggett-Garg inequality (LGI) violations.

Readout errors can be characterized using a calibration matrix \( A \) of size \( N \times N \), where \( N = 2^n \) and \( n \) is the number of qubits used in the experiment. The elements \( A_{ij} \) of this matrix represent the probability of obtaining the bit string \( j \) when the ideal output should have been \( i \). These probabilities are determined through calibration experiments on individual qubits, assuming that readout errors on each qubit are independent of those on other qubits. Using the calibration matrix \( A \), the ideal probabilities \( \vec{p}_{\text{ideal}} \) can be derived from the noisy probabilities \( \vec{p}_{\text{noisy}} \) obtained in the experiment using the following equation:

\begin{equation}
    \vec{p}_{\text{noisy}} = A \cdot \vec{p}_{\text{ideal}}
\end{equation}

The size of the calibration matrix \( A \) grows exponentially with the number of qubits \( n \). While this is not a significant issue for our experiment, it becomes a non-trivial challenge for experiments involving a large number of qubits. This is where the true power of Mthree becomes evident, as it is particularly effective for scaling to systems with many qubits.

We utilized the Mthree command \verb|M3Mitigation.cals_from_system()|
 to compute the calibration matrix for the qubits used in the experiment. Furthermore, we applied \verb|M3Mitigation.apply_correction()| to obtain the corrected probabilities. The experiment was repeated and Mthree error mitigation techniques were applied to generate the readout error-mitigated results, as illustrated in Figure \ref{fig:mthree}. As shown in the appendix, readout errors systematically reduce the LGI violation values below the expected levels. The application of readout error mitigation significantly improved these values, bringing them closer to the theoretically expected results. The Root Mean Square Error (RMSE), calculated by squaring the difference between the experimental results and expected values, then averaging over all experiments, is $0.00073$ without error mitigation. This improves to $0.000183$ after applying error mitigation.

Currently, the Sampler does not have the capability to mitigate gate errors, which were a minor source of error in our experiment. However, this can be addressed in the future as such methods are adopted to enhance result precision. 
\begin{figure}
    \centering
    \includegraphics[width=1\linewidth]{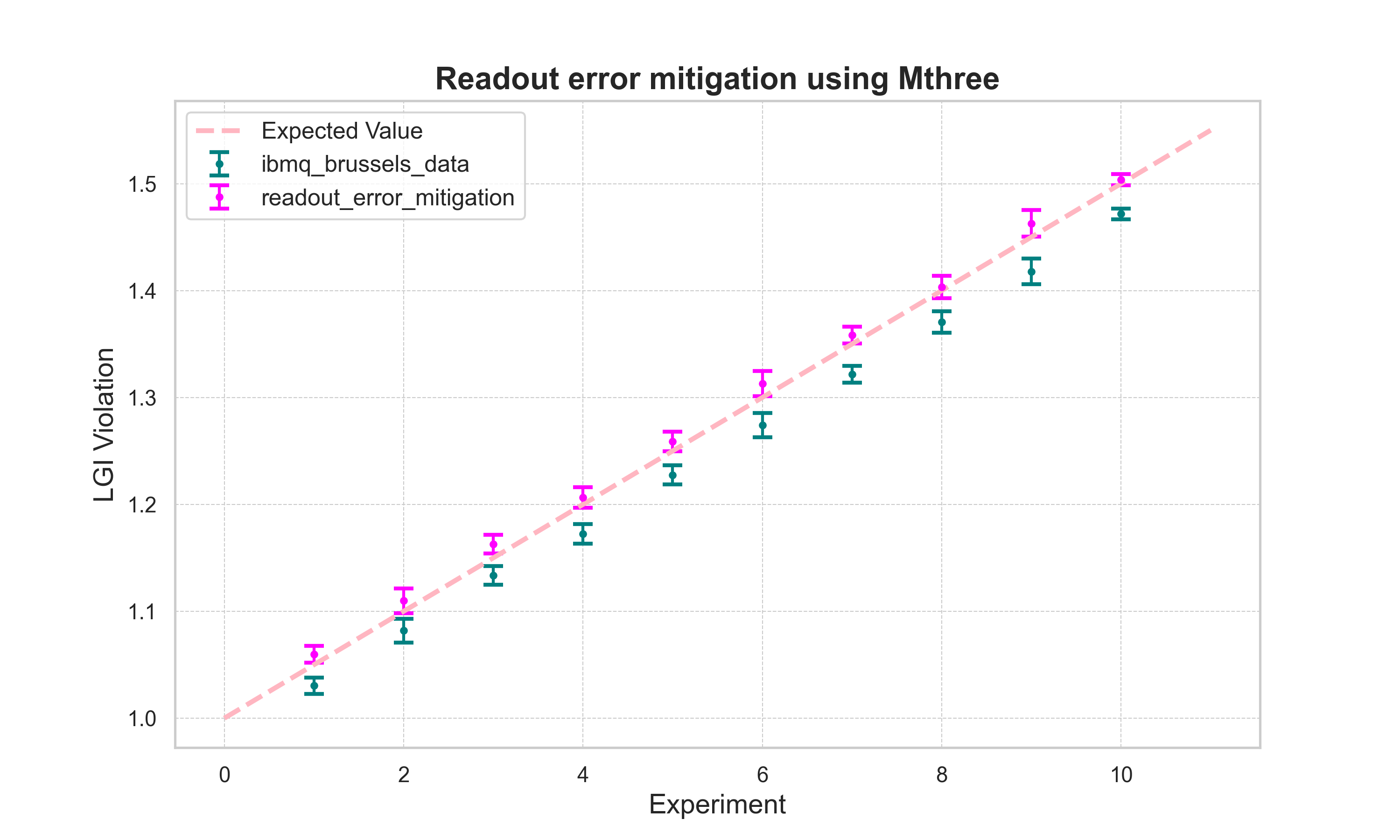}
    \caption{Comparison of raw and readout error-mitigated values of LGI violations, performed in IBM Brussels. For each LGI violation, the experiment was repeated \(10\) times, with \(20,\!000\) shots per experiment. The readout error-mitigated values, obtained using Mthree's correction techniques, are elevated and align more closely with the expected theoretical values, demonstrating the effectiveness of the mitigation process.}
    \label{fig:mthree}
\end{figure}

\section{Qiskit: Advanced Functions and Challenges}
During the final stages of our experiment, we utilized advanced functionalities of the latest version of Qiskit, such as the \textit{Sampler} and \textit{Batch} features. These tools proved to be highly effective in implementing error mitigation strategies, significantly enhancing the reliability of our results. Although most of our outcomes aligned well with theoretical expectations, we occasionally observed results that were inconsistent or uncorrelated with the expected behavior. These anomalies, though infrequent, highlight the inherent challenges and variability associated with current quantum computing hardware. Despite these occasional discrepancies, the advanced capabilities of Qiskit provided a robust framework for achieving meaningful and reproducible results in our study.

\section{Conclusion}

We generated secure random numbers certified by the principles of Quantum Mechanics, utilizing IBMQ backends, specifically  \textit{Brussels}, \textit{Perth},  \textit{Lagos}, and \textit{Kyoto}. Certification of these random numbers was achieved through the successful violation of the Leggett-Garg Inequality and compliance with the No Signaling in Time conditions. The implemented protocol is notably simple, requiring minimal circuits composed of gates that can be executed with high accuracy and minimal errors. Additionally, we conducted a thorough noise analysis to demonstrate and understand the impact of noise on our experimental outcomes.

It is important to note that this implementation is not devoid of loopholes. Enhancements can be made to conduct a completely loophole-free experiment, thereby fortifying the Randomness Generation Protocol. As this process is conducted in the cloud, sub-runs are performed one after another without specifying a seed. Alternatively, incorporating a random seed and implementing an extraction procedure can further secure the generated bits. Random numbers were generated using a random seed in the qiskit simulator as shown in Appendix.

This work also serves as a fundamental validation of quantum mechanics on a quantum computer. In addition to contributing to a growing body of quantum mechanical tests\cite{sadana2022testing,sadana2023noise,santini2022experimental} conducted on quantum computers, it also has practical applications for benchmarking quantum devices. Given that our test requires only a single qubit, it provides a straightforward method for benchmarking individual qubits as well.

Although attempts were made to generate random numbers using a quantum computer \cite{li2021quantum, jacak2021quantum, orts2023quantum} none of them provide certification which is required for secure random number generators. Furthermore, conducting a Bell experiment on cloud-based quantum computers is not feasible due to the difficulty in satisfying the no-signaling condition, which is essential for certifying randomness. In contrast, for temporal scenarios, the No Signaling in Time condition does not present any obstacles to implementation. This proof-of-concept demonstration, which showcases the generation of secure random numbers using cloud-based quantum computers, is anticipated to stimulate further research. The objective is to develop more robust protocols suitable for implementation in commercial-grade quantum computers when they become accessible to the general public.

\textit{Acknowledgments : }We especially thank Dipankar Home for useful discussions. We would like to extend our sincere gratitude to Sean Wagner(IBM) for his invaluable assistance in utilizing the advanced functionalities of Qiskit. His expertise significantly contributed to improving the quality and accuracy of our results. We are also deeply grateful to Jagan Natarajan(IBM) for his guidance and support in migrating our code to the newer version of qiskit.   U.S. acknowledges partial support provided by the Ministry of Electronics and Information Technology (MeitY), Government of India under a grant for Centre for Excellence in Quantum Technologies with Ref. No. 4(7)/2020-ITEA, partial support from the QuEST-DST Project Q-97 of the Government of India as well as from a Canada Excellence Research Chair professorship. AS acknowledges support from SERB core grant CRG/2021/000873 and from a Quantum Horizons Alberta chair professorship. We acknowledge the use of IBM Quantum Credits for this work. The views expressed are those of the authors, and do not reflect the official policy or position of IBM or the IBM Quantum team.

\bibliography{references}

\clearpage
\newpage
\onecolumngrid
{\begin{center}\bf \Large{Supplementary material}\end{center} }

\begin{figure}[H]
    \centering
    \includegraphics[width=1\linewidth]{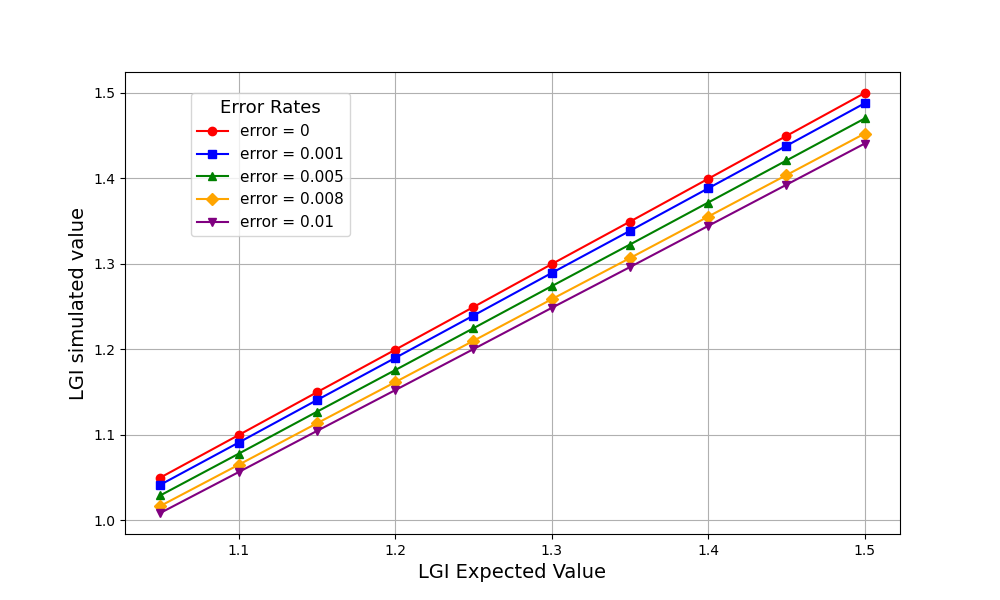}
    \caption{Numerical error analysis of the experiment using $Z$ gates applied once before each measurement with varying probabilities $p$. The observed value of the Leggett-Garg inequality (LGI) decreases with increasing probability of the $Z$ gates, indicating that measurement errors are a source of error in the experiment.}
    \label{fig:error_analysis}
\end{figure}

\textbf{\textit{Analytical Noise Analysis}}: In order to further isolate the noise, we did a theoretical analysis on which type of noise affected our result. We added $X$ operators with a certain probability in the calculation of the joint probabilities. 
We then added $Z$ operators in our calculations with a probability of $p$ and observed that the expected LGI value decreases as we increase the probability $p$ as shown in figure \ref{fig:error_analysis}

\textbf{\textit{LGI violation using random seeds:}} To demonstrate a thorough analysis of how our algorithm can be enhanced to generate random numbers on a quantum computer, we used a random seed to select one of the three circuits employed to compute the three two-time correlations, as shown in Figure \ref{fig: Circuit post transpilation}. Each time, the seed determines the specific circuit to be run, and the results are then compiled to compute the LGI value. This step is computationally expensive on a quantum computer because it requires running a different circuit each time, so we used the Qiskit simulator for these experiments. The simulator, being free from noise, served as a preliminary verification step before executing the results on a real quantum computer.


\begin{figure}[H]
    \centering
    \begin{minipage}{0.32\textwidth}
        \centering
        \includegraphics[width=\textwidth]{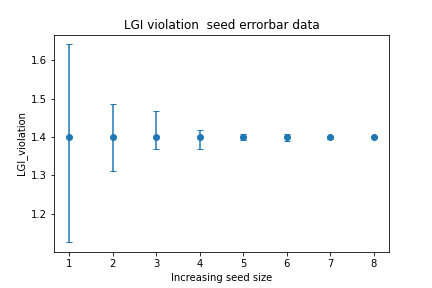}
    \end{minipage}
    \begin{minipage}{0.32\textwidth}
        \centering
        \includegraphics[width=\textwidth]{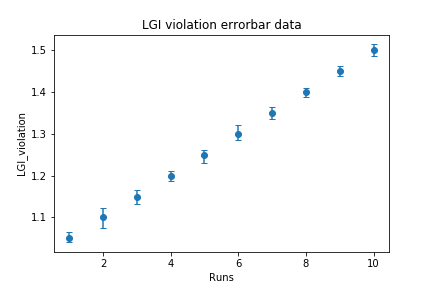}
    \end{minipage}
    \begin{minipage}{0.32\textwidth}
        \centering
        \includegraphics[width=\textwidth]{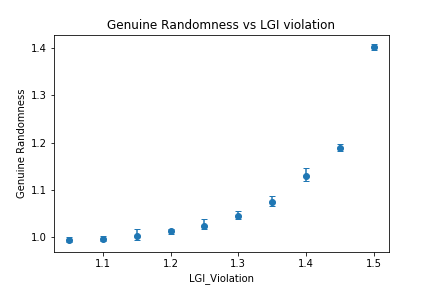}

    \end{minipage}
    \caption{LGI violation for varying seed ranging from $10^1$ to $10^8$ for the LGI violation of 1.4. The experiment was repeated $20$ times for a particular value of the seed and we see that the spread merges with the actual value as we increase the size of the seed.
    LGI violation data for the seed of length $10^5$. We repeated each experiment for a certain LGI value $10$ times using different seeds of the same length and plotted the spread. The corresponding genuine randomness value computed from the probabilities of these experiments are plotted corresponding to the LGI violation and it follows the lower bound on Genuine Randomness as derived in \cite{nath2024single}}
    \label{fig: LGI violation qiskit}
\end{figure}

We conducted experiments with seed sizes ranging from $10^1$ to $10^8$ for an LGI violation of 1.4 and observed that as the seed size increases, the spread of values decreases, and the LGI violation results converge more closely to the expected value(Figure \ref{fig: LGI violation qiskit}). We then extended the experiment to multiple LGI values using a seed size of $10^5$, finding that the results closely align with the expected values(Figure \ref{fig: LGI violation qiskit}). Additionally, we computed the genuine randomness based on the observed probabilities and found that it closely adheres to the randomness bound derived in \cite{nath2024single}, thus verifying the lower bound.


\textbf{\textit{Other backends:}}: Apart from our main experiment at IBMQ Brussels, we initially implemented our algorithm using IBMQ's free access backends, specifically IBMQ Perth, IBMQ Lagos, and IBMQ Kyoto. While these backends were noisier compared to the main backend, they still produced fairly decent results, as shown in Figure \ref{fig:main}. This indicates that our protocol can be effectively implemented even without fully noise-mitigated backends.

\begin{figure}
    \centering
    \begin{minipage}{0.32\textwidth}
        \centering
        \includegraphics[width=\textwidth]{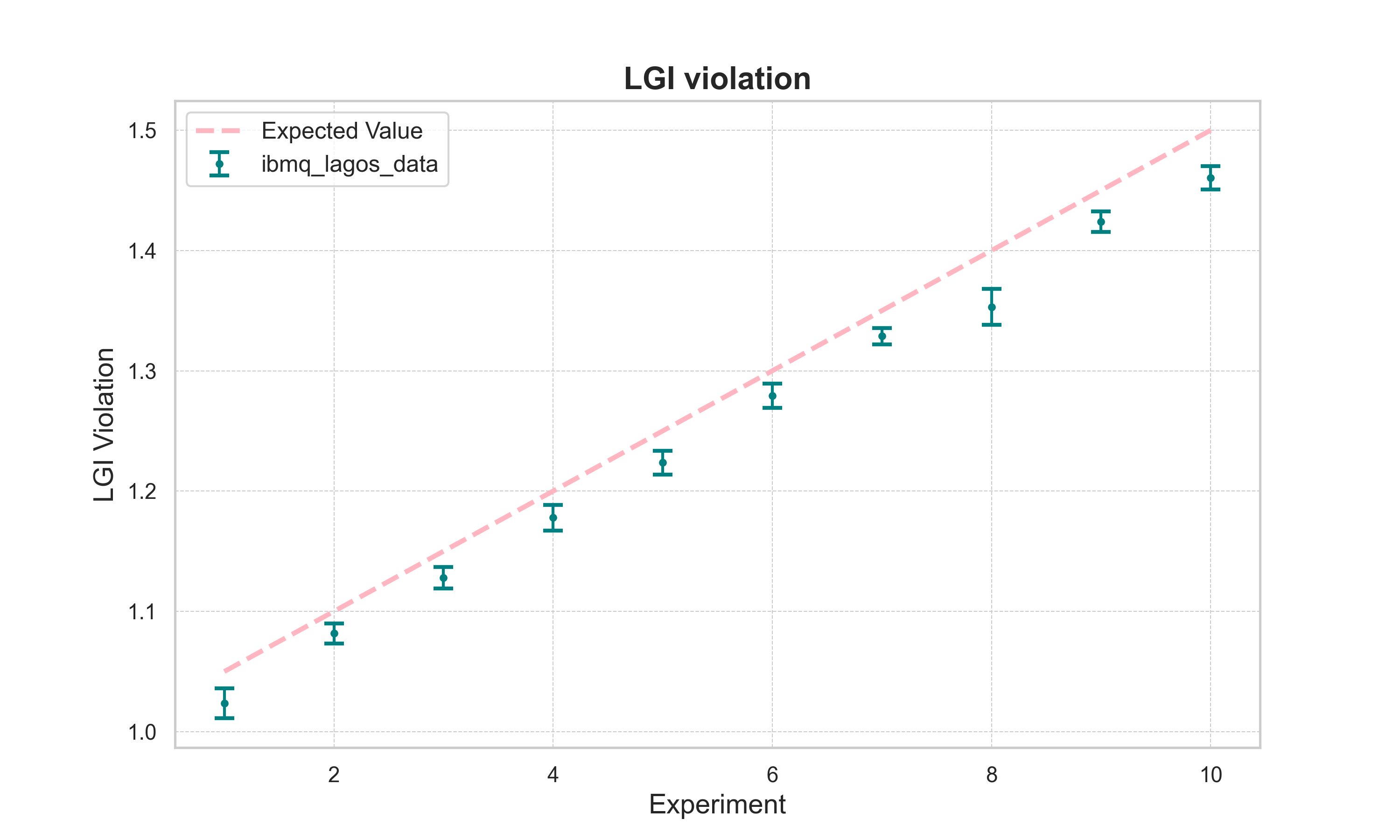}

    \end{minipage}
    \begin{minipage}{0.32\textwidth}
        \centering
        \includegraphics[width=\textwidth]{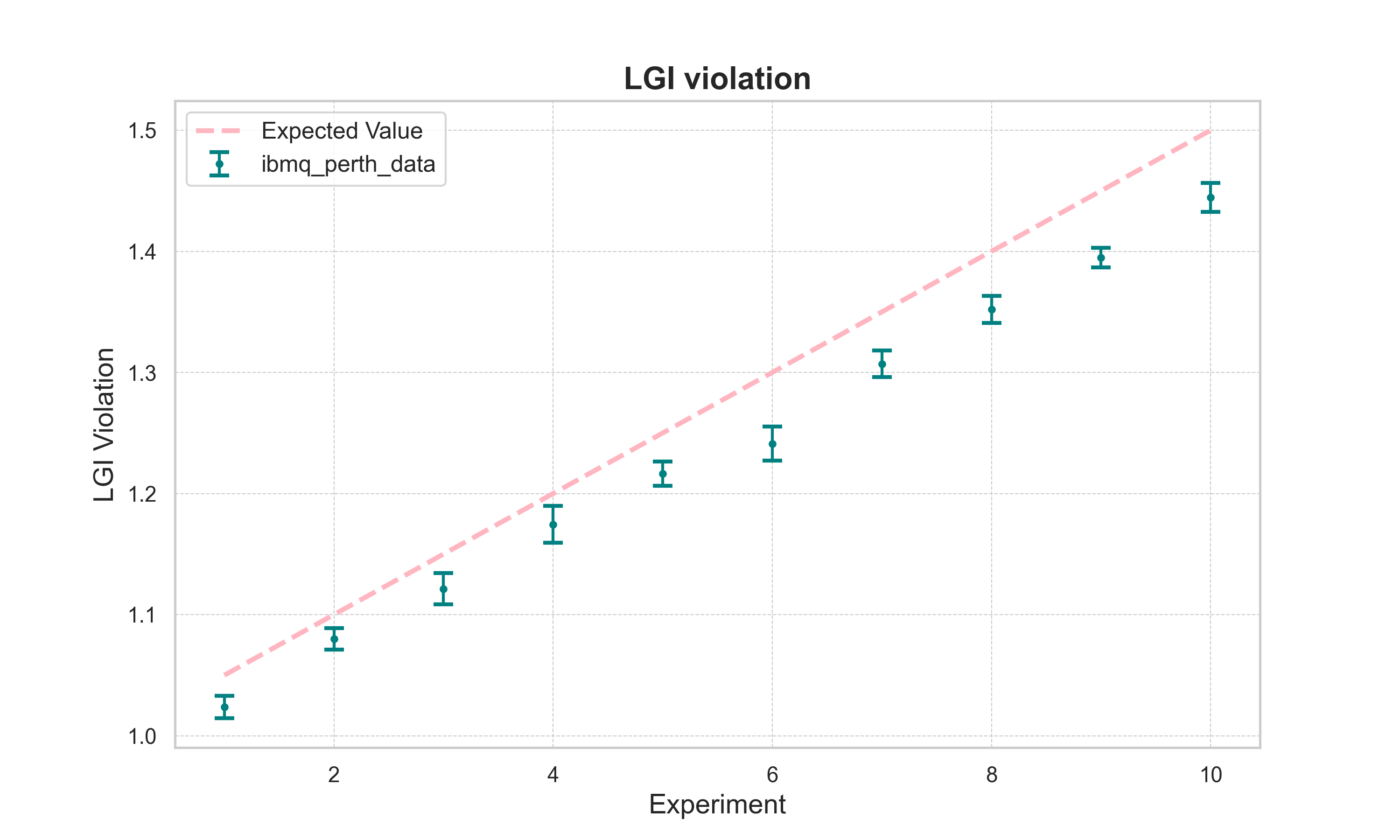}
    \end{minipage}
    \begin{minipage}{0.32\textwidth}
        \centering
        \includegraphics[width=\textwidth]{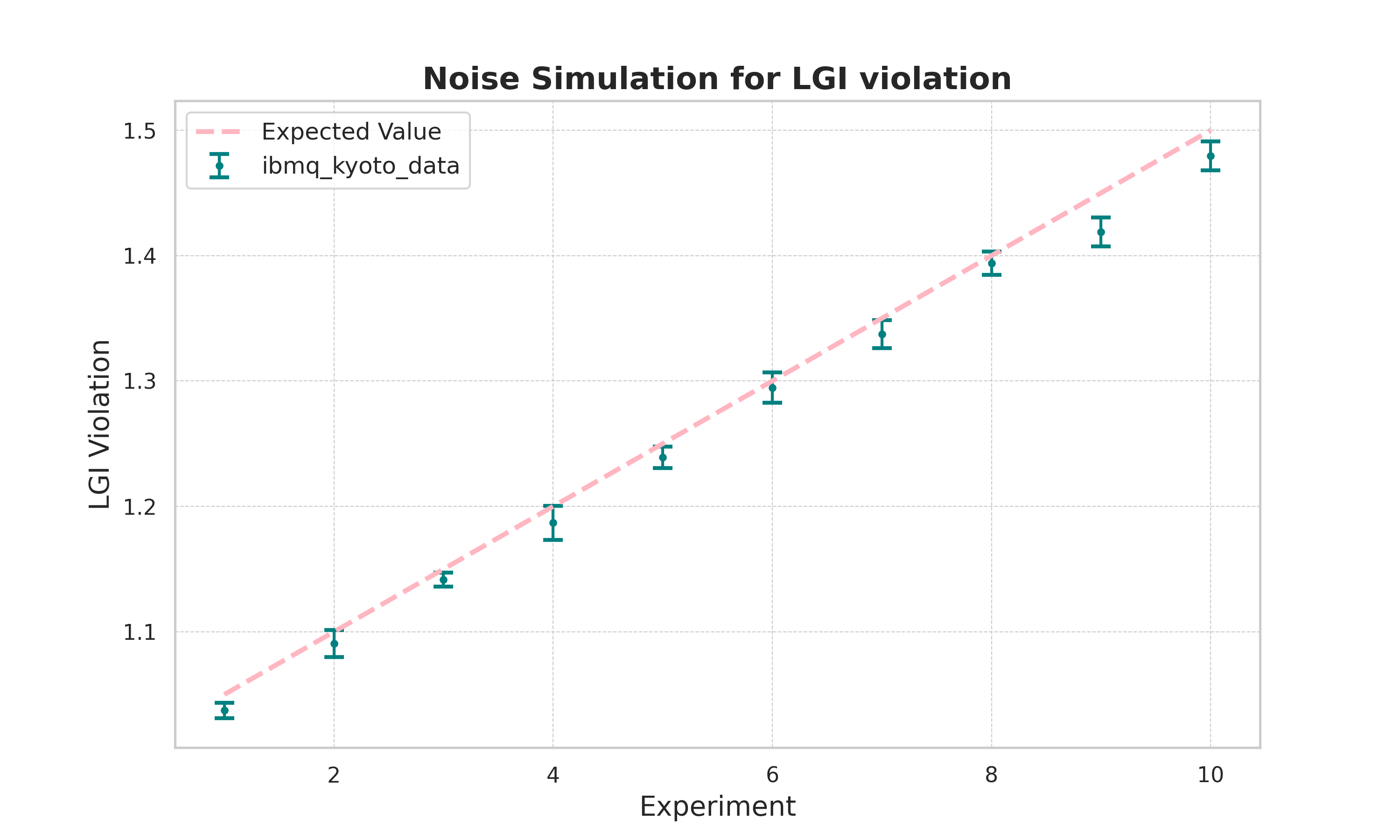}

    \end{minipage}
    \caption{LGI violation for IBM Lagos, IBM Perth, and IBM Kyoto respectively.  Error Bar of the results vs. Expected LGI violation are obtained on IBM hardware over 10 runs of the experiment.}
\end{figure}



\begin{figure}[H]
    \centering
    \begin{minipage}{0.32\textwidth}
        \centering
        \includegraphics[width=\textwidth]{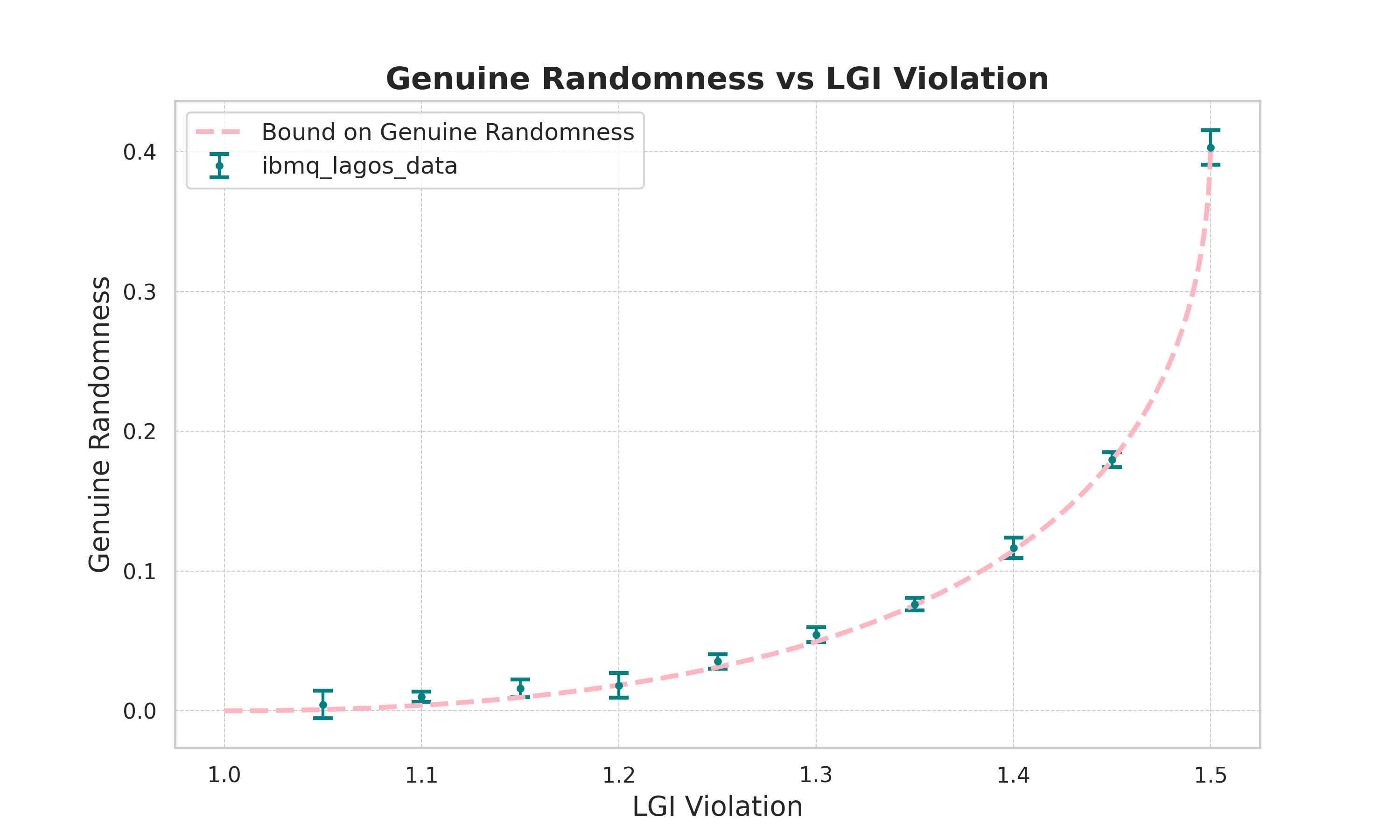}

    \end{minipage}
    \begin{minipage}{0.32\textwidth}
        \centering
        \includegraphics[width=\textwidth]{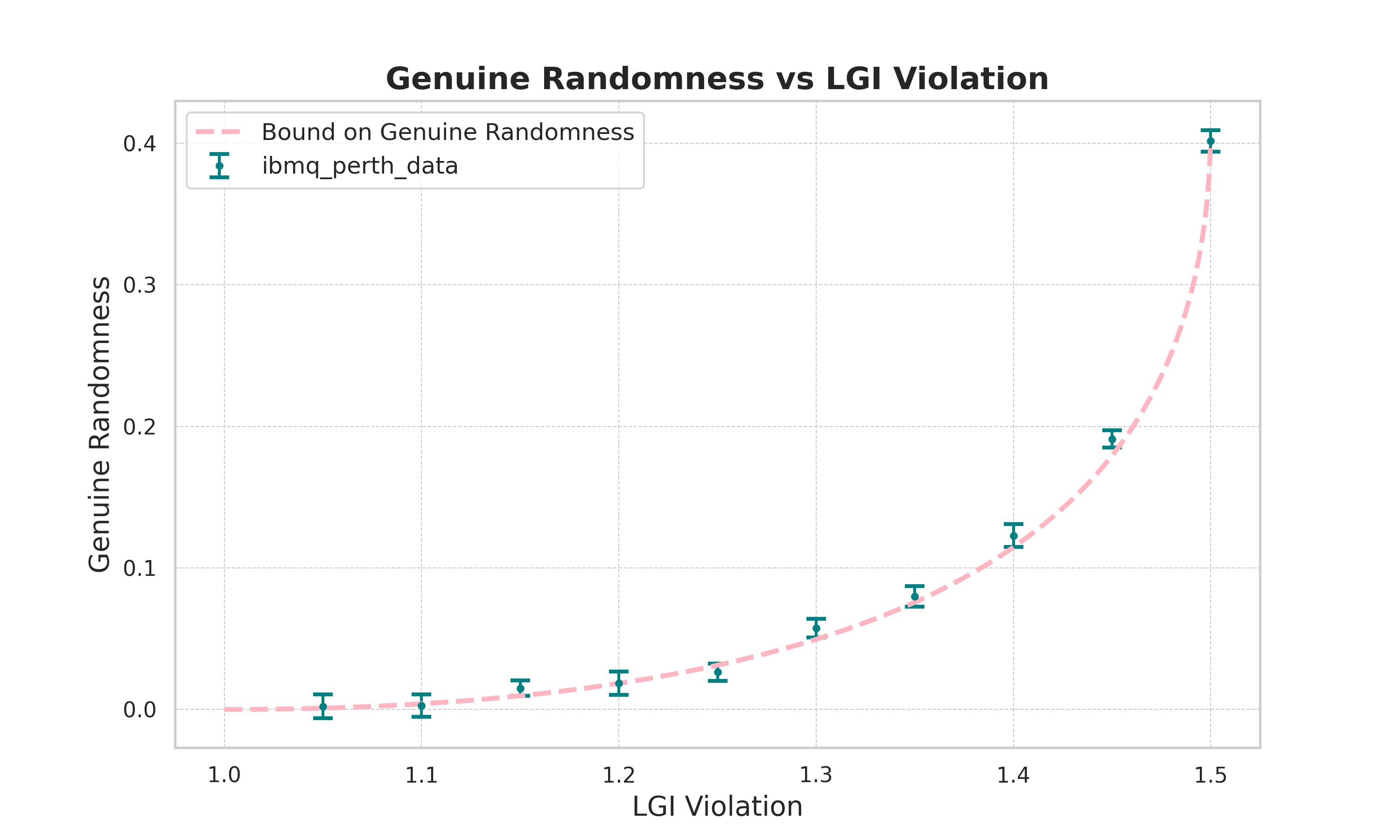}
    \end{minipage}
    \begin{minipage}{0.32\textwidth}
        \centering
        \includegraphics[width=\textwidth]{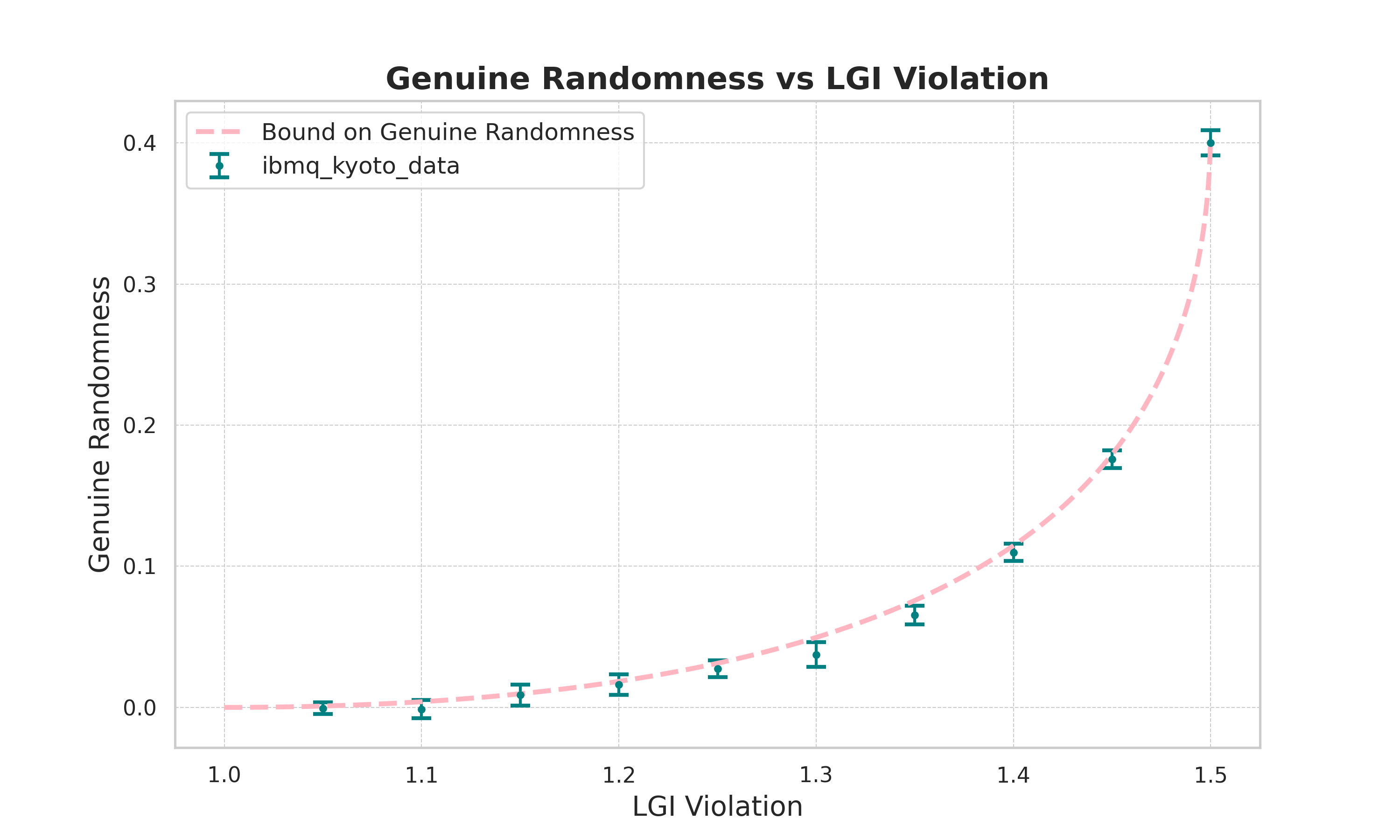}

    \end{minipage}
    \caption{Genuine Randomness for IBM Lagos, IBM Perth, and IBM Kyoto respectively.  Error Bar of Genuine Randomness vs. LGI violation are obtained on IBM hardware over 10 runs of the experiment.}
\end{figure}

\textbf{\textit{Random Number Generation using mixed states}}: We can obtain the protocol for randomness generation for any LGI value by using the following mixed state(Eq \ref{eq:mixed_state}), a combination of the two pure states $\ket{\psi_1} = \ket{0}$ and $\ket{\psi_2} = \ket{1}$. The parameters in the circuit for certain LGI violation using mixed states is shown in the table \ref{table:mixed states}.

\begin{equation}
    \rho = \frac{1}{2}\ket{\psi_1}\bra{\psi_1} + \frac{1}{2}\ket{\psi_2}\bra{\psi_2} 
    \label{eq:mixed_state}
\end{equation}
\vspace{10 mm}

\begin{table}[H]
\label{mixed state params table}
\centering
\renewcommand{\arraystretch}{1.3} 
\setlength{\tabcolsep}{12pt} 
\begin{tabular}{|l|l|l|}
\hline
LGI & $\theta_1$ & $\theta_2$  \\
\hline
1.05 & 6.25752 & 11.8 \\
\hline
1.10 & 6.23037 & 11.8 \\
\hline
1.15 & 6.20133 & 11.8 \\
\hline
1.20 & 6.16983 & 11.8 \\
\hline
1.25 & 6.13493 & 11.8 \\
\hline
1.30 & 6.09496 & 11.8 \\
\hline
1.35 & 6.04625 & 11.8 \\
\hline
1.40 & 5.97623 & 11.8\\
\hline
1.45 & -101.212 & 128.279 \\
\hline
1.50 & 147.131 & -48.6475 \\
\hline
\end{tabular}%

\caption{Values of the parameters $\theta_1$ and $\theta_2$ for the corresponding LGI violation starting with a mixed state.}
\label{table:mixed states}
\end{table}

\end{document}